\documentclass[lettersize,journal]{IEEEtran}
\usepackage[utf8]{inputenc}
\usepackage[T1]{fontenc}

\usepackage{amsmath,amssymb,amsfonts}
\usepackage{hyperref}
\usepackage{float}
\usepackage{tikz}
\usepackage[shortlabels]{enumitem}
\usepackage{array}
\usepackage{booktabs}
\usepackage[shrink=10,stretch=10]{microtype}
\usetikzlibrary{calc}

\newtheorem{defi}{Definition}
\newtheorem{prop}{Proposition}
\newtheorem{lemma}{Lemma}
\newtheorem{remark}{Remark}
\newtheorem{example}{Example}

\newcommand{\share}[1]{[#1]}

\newcounter{protocolcounter}

\newfloat{protocolFloat}{tb}{pFloat}
\newcommand{\protocolEnum}{\begin{enumerate}[1.,leftmargin=*]}
  \makeatletter
  \newcommand{\protName}{Protocol}
  \makeatother
  \newenvironment{protocol}[3][tb]{%
      \refstepcounter{protocolcounter}\unskip\ignorespaces
      \setlist[enumerate]{label=(\alph*)}
      \begin{protocolFloat*}[#1]
        \begin{minipage}{\textwidth}
          \small
          \hrule height 1.5pt
          \vglue 5pt
          {\centering{\bfseries\protName~\theprotocolcounter: ~#2}\phantomsection\par}
          \unskip
          \vglue 5pt
          #3\par\vglue 5pt\hrule\par\vglue 5pt
          \setlist{
              topsep=0pt,
              itemsep=-0.5pt,
          }
          \protocolEnum
      }
      {
        \end{enumerate}
      \end{minipage}
      \vglue 4pt
      \hrule height 1.5pt
    \end{protocolFloat*}
    \unskip\ignorespacesafterend
}

\newcommand{\orcidID}[1]{
    \unskip\hspace{0.1em}\href{https://orcid.org/#1}{\raisebox{-0.12\height}{\includegraphics[height=0.8em]{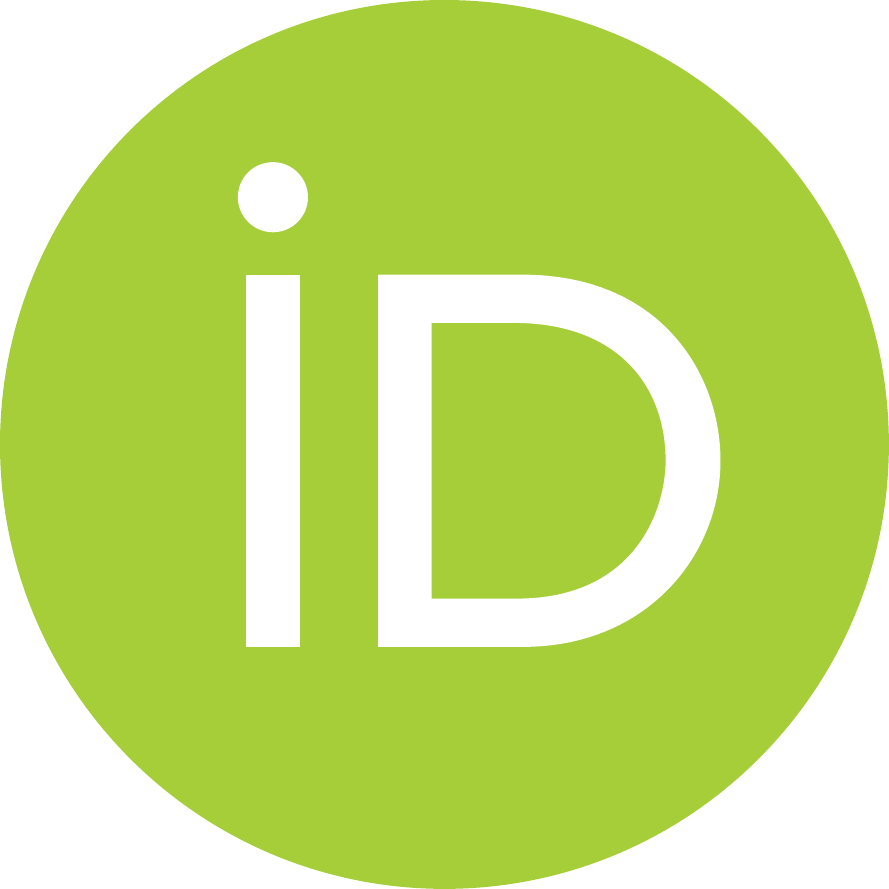}}}%
}

\begin{document}

\title{Semi-Private Computation of Data Similarity with Applications to Data Valuation and Pricing}
\author{René~Bødker~Christensen\orcidID{0000-0002-9209-3739},
    Shashi Raj Pandey\orcidID{0000-0002-5781-4131},
    and Petar Popovski\orcidID{0000-0001-6195-4797}%
    \thanks{This paper was supported by the Villum Investigator Grant ``WATER'' from the Velux Foundations, Denmark.}%
    \thanks{R.B. Christensen is with the Department of Mathematical Sciences, Aalborg University, and Department of Electronic Systems, Aalborg University.}%
    \thanks{S.R. Pandey and P. Popovski are with the Department of Electronic Systems, Aalborg University.}}

\markboth{Article Postprint}%
{Christensen \MakeLowercase{\textit{et al.}}: Semi-Private Computation of Data Similarity}

\IEEEpubid{%
    \begin{minipage}{\textwidth}
      \vglue 1cm
      \centering
      ©2023 IEEE. Personal use of this material is permitted. Permission from IEEE must be obtained for all other uses, in any current or future media, including reprinting/republishing this material for advertising or promotional purposes, creating new collective works, for resale or redistribution to servers or lists, or reuse of any copyrighted component of this work in other works. DOI: \href{https://doi.org/10.1109/TIFS.2023.3259879}{10.1109/TIFS.2023.3259879}
    \end{minipage}%
}

\maketitle

\begin{abstract}
  Consider two data providers that want to contribute data to a certain learning model. Recent works have shown that the value of the data of one of the providers is dependent on the similarity with the data owned by the other provider. It would thus be beneficial if the two providers can calculate the similarity of their data, while keeping the actual data private. In this work, we devise multiparty computation-protocols to compute similarity of two data sets based on correlation, while offering controllable privacy guarantees. We consider a simple model with two participating providers and develop methods to compute exact and approximate correlation, respectively, with controlled information leakage. Both protocols have computational and communication complexities that are linear in the number of data samples. We also provide general bounds on the maximal error in the approximation case, and analyse the resulting errors for practical parameter choices.
\end{abstract}


{\section{Introduction}}
\subsection{Background and Motivation}\label{subsec:background}
\IEEEPARstart{W}{ith} an increased roll-out of networked cyber-physical systems such as Internet of Things (IoT), there has been an unprecedented growth in collection and analysis of data. This development has also given rise to data markets \cite{liang2018survey, acemoglu2019too, agarwal2019marketplace}, where data is exchanged between different entities.
For instance, large quantities of high-quality data is fundamental during the training of machine learning models, and this may require the acquisition of data from external data holders. As a specific example, \cite{pinson2022regression} considers a regression market, where a forecasting model is trained based on data distributed among different data providers such as IoT-devices.
That is, the data providers -- or simply referred to as \emph{participants} -- perform collaborative training in a market setup to improve forecasts.

There are, however, several challenging aspects in making such a market work. One such challenge is the fact that participants holding `similar' data -- in particular, correlated data -- will be able to contribute similar knowledge to the model training. Once data from one participant is known, however, the data held by the other participant will have limited value to whoever is training the model \cite{agarwal2019marketplace, ghorbani2019data}.
This may lead to \emph{data rivalry} \cite{acemoglu2019too, agarwal2019marketplace}, where participants are incentivized to sell earlier than others, potentially underbidding each other and thereby decreasing the overall price of the data. This injects mistrust in the data market and participants may opt out from exchanging their data at all.

\IEEEpubidadjcol
With this in mind, it may be beneficial for participants to cooperate, sell the data collectively, and share the payment between them in the case where their data sets are correlated \cite{pandey2022participation}. In practice, though, these correlations are unknown a priori, and the participants are likely reluctant to reveal their data for computing correlations in the clear. Hence, there is a need for participants to determine if their data sets are similar without actually revealing their data sets to each other. A specific measure of similarity is the correlation\footnote{Correlation is an intuitive metric adopted widely to quantify data similarity in various data market designs \cite{agarwal2019marketplace,acemoglu2019too,pandey2022participation}}, so in other words, the participants aim to compute correlation in a private fashion.

Several studies on incentive mechanism design have been conducted to handle the behaviour of strategic participants \cite{fudenberg1991game, roth2000game}, elicit data from them while manipulating the privacy costs, and for different contexts, derive collaboration opportunities for accurate computation in the data markets, such as in regression markets. Nevertheless, such techniques demand participants to reveal parts of their private information. In fact, the loss of privacy in data markets leads directly to monetary loss. This also highlights the tension between data privacy and the benefits of data.

From these observation, we see that knowing correlation -- an important measure of data similarity -- between data samples of the participants helps to explore the following outlined key research question:
When it is beneficial for the participants to perform collaborative training? In other words, this can be stated as the problem of learning whether/when to collaborate or not.

In order to solve this question in general, one must first be able to solve it in the simplest case. That is, the case where each participant holds several samples of the same univariate random variable. This is our focus in this work, and the techniques presented here can later be extended to the general case as sketched towards the end of this work. Note that only the number of random variables is limited; the number of known samples can still be arbitrarily large.
Solving the univariate case reduces to computing a scalar product between privately held vectors, which is an interesting problem in its own right. Throughout, we keep the focus on the correlation coefficient, however, merely mentioning that the same techniques are applicable in more general problems.

Secure multiparty computation (MPC) is a term used to describe techniques that allow multiple participants to give inputs to and compute the value of a publicly known function without revealing the individual inputs. Traditionally, the functions under considerations in these MPC-problems are defined over a finite field.
In many real-world problems, however, the inputs of each participant will not be elements of a finite field but instead real numbers. The literature already contains a number of suggestions for dealing with such cases. One way is to use MPC techniques designed for floating-point numbers \cite{Aliasgari13,Kamm2015}. These methods are typically very computationally expensive, however. Another way is to assume that the inputs can be represented as fixed-point numbers, which can readily be mapped to a finite field where standard MPC techniques can be applied \cite{Catrina2010,Cock2015,Gascon17}.

Nevertheless, certain functions are inherently real in the sense that even under the assumption that the inputs are exactly representable as fixed-point numbers, the output cannot be guaranteed to be representable as a fixed-point number.
One such function is the sample correlation between two sets of data samples as in the data market example mentioned previously.

When exact computation with fixed-point numbers is not possible, we can in principle still use essentially the same approach by approximating the inputs by fixed-point numbers and obtaining the function value up to some approximation error. While this is an easy solution, it complicates security arguments as it will be necessary to prove that the error does not reveal more about the inputs than the desired function value~\cite{Feigenbaum2006}.

There is another reason one might prefer to work directly with real inputs rather than its implementation in hardware.
Namely, for many types of data the natural representation would be to use real numbers, and ideally, we would perform all computations using reals.
However, since it is not possible to perfectly represent arbitrary reals in practical implementations, a particular number system will be used to model the structure of the real numbers.
This representation could for instance be fixed-point numbers, IEEE~754 floating-point numbers~\cite{Ieee754}, BFloats~\cite{bfloat}, or posits~\cite{Gustafson17}.
These make different compromises in approximately capturing the behaviour of reals in a finite number of bits, and thus behave slightly differently.
In other words, the chosen representation is simply a consequence of the available hardware rather than an indication of the `ideal problem' that we want to solve.
By analysing the problem under the assumption that inputs are real, the results we obtain are not tied to specific types of hardware, current or future.

\subsection{Related works}\label{subsec:related-work}
There are numerous works that treat privacy-preserving computation on real or floating-point numbers. These take different approaches as some work directly with real numbers \cite{Tjell21}, some with floating-point numbers as specified by the IEEE 754 standard \cite{Pullonen15}, and yet others using alternative number representations \cite{Dimitrov16,Kamm2015,Liu16}.
There has also been an interest in finding the best way to quantify the amount of leakage for protocols using real numbers~\cite{Kuskonmaz22}.

An alternative approach to the problem at hand would be to consider differential privacy~\cite{Dwork06,Dwork08}, which would ensure that the output would essentially remain the same if a single sample point were to be left out. This thus provides privacy for the individual data samples, while allowing information about the full data set to be extracted.
Note, however, that from the point of view of the data provider in our setting, the value lies in the distribution of the full data set, with the individual data point providing little value in itself. Hence, our goal is somewhat different to the one provided by differential privacy. It has, nonetheless, been used in training of machine learning models -- see e.g. \cite{Pentyala22,Gong20} -- which is a go-to setting to consider for a data market.

Following \cite{acemoglu2019too}, a series of works \cite{agarwal2019marketplace,ali2020voluntary,jones2020nonrivalry} have explored the combinatorial nature of data, highlighting the negative effect of sharing correlated data on pricing and market efficiency. It was formally proven that correlated data can cause data trading problems \cite{acemoglu2019too, ali2020voluntary}; such information leakage leads to price depression and uncontrolled data exchanges, leaving poor communication and training efficiency during decentralized model training. A way out envisioned to avoid this unwanted scenario would be to -- a priori -- privately know data correlation between participants and accordingly develop the data revealing strategies.
Yet, a majority of works~\cite{acemoglu2019too, agarwal2019marketplace, cummings2015truthful} focus on designing market mechanisms for data acquisition in an attempt to incentivize truthful participation. This leaves the effect of correlation in data trading open. Several game-theoretic approaches or auction mechanisms \cite{agarwal2019marketplace, feng2021uncovering, pandey2020crowdsourcing} are employed to formulate such interactions between the data market and the participants with some privacy guarantees, trading differentially private data samples. Nevertheless, there is a gap in the literature on designing efficient protocols to privately compute sample correlation, either on real or floating-point numbers. These protocols are fundamental for realizing an efficient market design. 

\subsection{Our Contributions}\label{subsec:contributions}
In this work, we focus on computing the sample correlation in a way that is efficient, but still retains reasonable privacy guarantees. More precisely, we describe a protocol that allows the participants to approximate the sample correlation to a high degree of accuracy, and then show that this is at least as secure as a protocol where the participants reveal certain values related to the rounding errors when representing their inputs as fixed-point numbers. In this way, we can explicitly describe those values that the participants should accept leakage of (as opposed to performing the computation on approximated fixed-point values without considering what might be revealed by the potentially systematic approximation errors).

In both of the protocols we present, the main computation is essentially a scalar product between two vectors, each one held locally by the corresponding participant. While such scalar products have previously appeared in the MPC literature, see e.g.~\cite{Gascon17,Luo09,Hu16}, the existing security analyses have---to the best of our knowledge---been based on the assumption that inputs are either fixed- or floating-point numbers rather than arbitrary reals. As pointed out in~\cite{Luo09}, one can find works which claim to work for real numbers, but are in fact insecure.

The main contribution of this work includes:
\begin{itemize}
    \item A general multi-party computation protocol to compute similarity of two data sets based on sample correlation is proposed.
    \item Two methods are proposed to compute sample correlation with reasonable privacy leakage.
    \item Rigorous analysis is done to evaluate the computation and communication complexities of the proposed methods. 
    \item We evaluate the magnitudes of approximation errors in practical implementations of the developed methods.
\end{itemize}

The rest of this paper is structured as follows. Section~\ref{sec:preliminaries} recalls the basics of MPC and fixed-point numbers that will be used in subsequent sections. It also clarifies our model assumptions.
In Section~\ref{sec:semi-priv-comp}, we consider an exact correlation protocol where the participants reveal certain error values, and Section~\ref{sec:appr-corr} treats the approximate protocol, whose leakage is bounded by the leakage in the exact protocol. Furthermore, it contains an analysis of the possible protocol parameters in a practical implementation.
Section~\ref{sec:tradeOff} contains a discussion of the trade-off between privacy and precision provided by the proposed protocols.
Finally, Section~\ref{sec:open-problems} lists some open problems, and Section~\ref{sec:conclusion} contains the concluding remarks.

\section{Preliminaries}\label{sec:preliminaries}
\subsection{Secure Multiparty Computation}\label{sec:mpc}
For a prime $p$, let $\mathbb{F}_p$ be the finite field of $p$ elements, and consider two participants $P_1$ and $P_2$, each with private inputs $\mathbf{x}^{(1)}\in\mathbb{F}_p^s$ and $\mathbf{x}^{(2)}\in\mathbb{F}_p^s$, respectively, where $\mathbb{F}_p^s$ denotes the $s$-fold Cartesian product $\mathbb{F}_p\times\cdots\times\mathbb{F}_p$. The goal of MPC is to devise a protocol for computing $f(\mathbf{x}^{(1)},\mathbf{x}^{(2)})$ that reveals no more about the inputs $\mathbf{x}^{(1)},\mathbf{x}^{(2)}$ than what is revealed by the value of $f(\mathbf{x}^{(1)},\mathbf{x}^{(2)})$ itself.

This latter condition is referred to as \emph{privacy}.
Initially, one might think that we should require that a corrupt participant learns \emph{nothing} about the input of the honest participant. But this is too restrictive since the value of $f(\mathbf{x}^{(1)},\mathbf{x}^{(2)})$ combined with the participants own input may reveal non-trivial information about the other input. As a simple example, computing the sum $f(x^{(1)},x^{(2)})=x^{(1)}+x^{(2)}$ will always allow each participant to learn the input of the other participant. Therefore, we should require that a corrupt participant learns \emph{no more than what is revealed by the function value itself}. The standard way to formalize this, is essentially to require that protocol transcripts in the \emph{real world} -- that is, when using the protocol -- are indistinguishable from an \emph{ideal world} where the function is computed by an ideal functionality and the remaining parts of the transcript are produced by a simulator~\cite{Canetti20}. One can think of the ideal functionality as a completely trusted third party, meaning that it is secure by definition.

The second requirement for an MPC-protocol is that it is \emph{correct}. Here, correctness refers to the protocol correctly implementing the functionality that it is meant to provide -- essentially that when $P_1$ and $P_2$ input $\mathbf{x}^{(1)}$ and $\mathbf{x}^{(2)}$, their output will be $f(\mathbf{x}^{(1)},\mathbf{x}^{(2)})$ as expected.

In greater detail, consider an ideal functionality $\mathcal{F}$ that given inputs $\mathbf{x}^{(1)}$ and $\mathbf{x}^{(2)}$ outputs $f(\mathbf{x}^{(1)},\mathbf{x}^{(2)})$ to both participants. In addition, let $\Pi$ be a protocol where the participants also output $f(\mathbf{x}^{(1)},\mathbf{x}^{(2)})$ when their inputs are $\mathbf{x}^{(1)}$ and $\mathbf{x}^{(2)}$ -- i.e. $\Pi$ computes $f$ correctly. Now, assume that $P_i$ is corrupted by the adversary. In the real world, $P_1$ and $P_2$ follow the protocol $\Pi$, and the adversary learns all that $P_i$ sees during the protocol execution. This `view' of the adversary is a random variable which we shall denote by $V_{\mathsf{real}}$.

In the ideal world, the honest participant, $P_{3-i}$ will give its input to $\mathcal{F}$, but the corrupted participant $P_i$ will be connected to a simulator $\mathcal{S}$ rather than $\mathcal{F}$. $P_i$ will then communicate with $\mathcal{S}$ and follow $\Pi$, with the messages from $P_{3-i}$ being simulated by $\mathcal{S}$. During this interaction, $\mathcal{S}$ should extract the input of $P_i$ and pass this onto $\mathcal{F}$. Again, the adversary sees whatever $P_i$ sees, which we will call $V_{\mathsf{ideal}}$.

To define privacy, we imagine a game where the adversary receives $V_{\mathsf{ideal}}$ or $V_{\mathsf{real}}$, and must distinguish whether it is one or the other. This must be difficult for $\Pi$ to be secure.
For the formal definition, we use distinguishers which take either $V_{\mathsf{ideal}}$ or $V_{\mathsf{real}}$ as input and produces a guess `$\mathsf{ideal}$' or `$\mathsf{real}$'.
\begin{defi}\label{defi:normalSecurity}
  Let $\mathcal{F}$ be an ideal functionality for computing $f$ and $\Pi$ a protocol that computes $f$ correctly.
  We say that $\Pi$ implements $\mathcal{F}$ with perfect security if
  \begin{equation*}
    \Pr[\mathcal{D}(V_{\mathsf{real}})=\mathsf{real}]=\Pr[\mathcal{D}(V_{\mathsf{ideal}})=\mathsf{real}]
  \end{equation*}
  for any distinguisher $\mathcal{D}$, where the probabilities are taken over the random choices of $P_1$ and $P_2$.
\end{defi}

One can think of Definition~\ref{defi:normalSecurity} as saying that whatever a participant sees during the protocol might as well be simulated by the participant itself, knowing only its own input and the output from the ideal functionality. Since the ideal functionality is secure by definition, this implies that $\Pi$ leaks no more than the ideal functionality.

For the protocols presented in this work, we will require correctness as usual, but relax the privacy requirements -- hence the `semi-private' in the title. What we mean by this is that we allow certain precisely defined values to leak during the protocols. Thus, we do not allow arbitrary leakage of a certain number of bits, instead stating explicitly the values that the participants allow to leak. We refer to this as \emph{controlled leakage} throughout.

More precisely, we will modify the real and ideal worlds in the following way, which is also illustrated in Figures~\ref{fig:realWorld} and \ref{fig:idealWorld}.
In the ideal world, we introduce a functionality $\mathcal{L}$, which we call a \emph{leaker}. This has an associated leakage function $\ell$, and before giving their inputs $\mathbf{x}^{(1)},\mathbf{x}^{(2)}$ to $\mathcal{F}$, they will send them to $\mathcal{L}$, which then leaks $\ell(\mathbf{x}^{(1)},\mathbf{x}^{(2)})$ to the simulator $\mathcal{S}$.
To match this in the real world, we introduce a null sink $\mathcal{N}$, which discards all inputs. Like in the ideal world, participants will send their inputs to $\mathcal{N}$ before starting the protocol.

\begin{figure}
  \centering
  \begin{tikzpicture}
    \tikzset{
        every node/.style={
            draw=gray!80,
            fill=gray!20,
            rectangle,
            minimum width=1cm, minimum height=1cm,
            rounded corners=1.5pt
        }
    }
    \draw(0,0) node(p1){$P_1$};
    \draw(3,0) node(p2){$P_2$};
    \draw(3,2) node(N){$\mathcal{N}$};

    \draw(p1.40) edge[->] (p2.140);
    \draw(p1.30) edge[<-] (p2.150);
    \draw(p1.15) edge[->] (p2.165);
    \draw(1.5,-0.05) node[draw=none, fill=none]{$\vdots$};
    \draw(p1.-40) edge[<-] (p2.220);

    \draw(p2.70) edge[->] (N.-70);
    \draw[->](p1) -- ++(0,1) -| (N.250);
  \end{tikzpicture}
  \caption{Illustration of the real world}
  \label{fig:realWorld}
  \bigskip
  \begin{tikzpicture}
    \tikzset{
        every node/.style={
            draw=gray!80,
            fill=gray!20,
            rectangle,
            minimum width=1cm, minimum height=1cm,
            rounded corners=1.5pt
        }
    }
    \draw(0,0) node(p1){$P_1$};
    \draw(3,0) node(p2){$P_2$};
    \draw(0,2) node(F){$\mathcal{F}$};
    \draw(3,2) node(L){$\mathcal{L}$};
    \draw(1.5,0) node(S){$\mathcal{S}$};

    \draw(p2.70) edge[->] (L.-70);

    \draw[->](S.120) |- (F.-20);
    \draw[<-](S.90) |- (F.20);
    \draw[->](L) -| (S.60);

    \draw[->](p1.60) -- ++(0,0.5) -| (L.250);
    \draw(p1.120) edge[->] (F.240);
    \draw(p1.90) edge[<-] (F.-90);

    \draw(S.40) edge[->] (p2.140);
    \draw(S.30) edge[<-] (p2.150);
    \draw(S.15) edge[->] (p2.165);
    \draw(2.25,-0.05) node[draw=none, fill=none]{$\vdots$};
    \draw(S.-40) edge[<-] (p2.220);
  \end{tikzpicture}
  \caption{Illustration of the ideal world}
  \label{fig:idealWorld}
\end{figure}

In a similar way to before, we denote by $V_{\mathsf{real}}^\ell$ and $V_{\mathsf{ideal}}^\ell$ the view of the adversary in the real and ideal worlds, respectively.
\begin{defi}
  Let $\mathcal{F}$ be an ideal functionality, $\mathcal{L}$ a leaker with leakage function $\ell$, and $\Pi$ a protocol. We say that $\Pi$ implements $\mathcal{F}$ with controlled leakage if
  \begin{equation*}
    \Pr[\mathcal{D}(V_{\mathsf{real}}^\ell)=\mathsf{real}]=\Pr[\mathcal{D}(V_{\mathsf{ideal}}^\ell)=\mathsf{real}]
  \end{equation*}
  for any distinguisher $\mathcal{D}$, where the probabilities are taken over the random choices of $P_1$ and $P_2$.
\end{defi}

Of course, this is a weaker security guarantee than given by Definition~\ref{defi:normalSecurity}. Note also that this privacy notion for a protocol $\Pi$ is tied to a specific choice of leakage function $\ell$. In particular, when using this definition later in Proposition~\ref{prop:protRevealErrorsSecure}, the leakage will describe the error terms revealed during Protocol~\ref{prot:revealErrors} defined in Section~\ref{sec:semi-priv-comp}.

An important tool in MPC is \emph{secret sharing}. In this work, we will use additive secret sharing, meaning that a secret $x\in\mathbb{F}_p$ is split into $n$ shares $(s_1,s_2,\ldots,s_n)\in\mathbb{F}_p^n$, where $s_i$ is chosen uniformly at random for $i=1,2,\ldots,n-1$ and $s_n=x-\sum_{i=1}^{n-1} s_i$. This is a secret sharing scheme with $(n-1)$-privacy and $n$-reconstruction. We use $\share{x}$ to denote a value shared in this way, and $\share{x}_i$ to denote the share held by the $i$'th participant.
More explicitly, writing $\share{x}$ in the case of two participants means that $P_1$ and $P_2$ hold values $\share{x}_1$ and $\share{x}_2$, respectively, such that $\share{x}_1+\share{x}_2=x$. 
For a thorough introduction to MPC, secret sharing, and proofs using simulators, see e.g. \cite{Cramer2015}.

We assume that the participants have access to a functionality $\mathcal{F}_\mathrm{BT}$ producing \emph{Beaver triples} \cite{Beaver91}, i.e., triples of shares $(\share{u},\share{v},\share{w})$ satisfying $uv=w$, where $u,v,w$ are in some finite field.
This allows the participants to compute products of secretly shared values. Namely, to compute the product of $\share{x}$ and $\share{y}$, they compute and reveal $x-u$ and $y-v$. It can then be checked that if the participants compute
\begin{align*}
  \share{z}_1 &=\share{w}_1+\share{u}_1(y-v)+\share{v}_1(x-u)+(x-u)(y-v)\\
  \share{z}_2 &=\share{w}_2+\share{u}_2(y-v)+\share{v}_2(x-u)
\end{align*}
using their respective shares, this yields $\share{z}_1+\share{z}_2=xy$. That is, $\{\share{z}_1,\share{z}_2\}$ is a sharing of $xy$.
In this way, the two participants can sacrifice a Beaver triple to multiply two shared values, and assuming that each triple is used only once, this multiplication is perfectly secure in theory. In practice, however, the security depends on the security of the method used to generate the triples.
Several such methods exist, including ones based on oblivious transfer \cite[Sec.\,4.1]{Gilboa99}\cite{MASCOT}, somewhat homomorphic encryption~\cite{Keller18}, and pseudorandom correlation generators~\cite{Boyle20}.

\subsection{Fixed-point representation}
\begin{figure*}[tb]
  \centering
  \newlength{\tickLen}
  \setlength{\tickLen}{1.5mm}
  \begin{tikzpicture}
    \tikzset{
        every edge/.style={draw=black,out=90, in=-90, out looseness=0.25, in looseness=0.6}
    }

    \draw(-6.5cm,0) node{$\mathbb{R}$};
    
    \draw[->, thick](-5.5cm,0) -- (-4.5cm,0) (-3cm,0) -- (2.5cm,0)  (4.5cm,0) -- (5.8cm,0);
    \draw[dashed, thick](-6cm,0) -- (-5.5cm,0) (-4.5cm,0) -- (-3cm,0) (2.5cm,0) -- (4.5cm,0);
    \draw[thick](-5,\tickLen) -- ++(0,-2\tickLen) +(0,-1mm) node(a)[anchor=north]{$-M\delta$};
    \draw[thick](-2,\tickLen) -- ++(0,-2\tickLen) +(0,-1mm) node(b)[anchor=north]{$-2\delta$};
    \draw[thick](-1,\tickLen) -- ++(0,-2\tickLen) +(0,-1mm) node(c)[anchor=north]{$-\delta$};
    \draw[thick](0,\tickLen) -- ++(0,-2\tickLen) +(0,-1mm) node(d)[anchor=north]{$0$};
    \draw[thick](1,\tickLen) -- ++(0,-2\tickLen) +(0,-1mm) node(e)[anchor=north]{$\delta$};
    \draw[thick](2,\tickLen) -- ++(0,-2\tickLen) +(0,-1mm) node(f)[anchor=north]{$2\delta$};
    \draw[thick](5,\tickLen) -- ++(0,-2\tickLen) +(0,-1mm) node(g)[anchor=north]{$M\delta$};

    \draw(0,-2cm) coordinate(collect) edge (a) edge (b) edge (c) edge (d) edge (e) edge (f) edge (g);
    \draw(0,-2.8cm) node(Q){$\mathbb{Q}_{(M,\delta)}$} edge (collect);

    \draw(2,-2.8) node{$\mathbb{F}_p$} edge[<-,in=0,out=180] node[above]{$\varphi_\delta$} (Q);
  \end{tikzpicture}
  \caption{Illustration of the fixed-point numbers in $\mathbb{Q}_{(M,\delta)}$ within the real numbers.}
  \label{fig:fixedPoint}
\end{figure*}
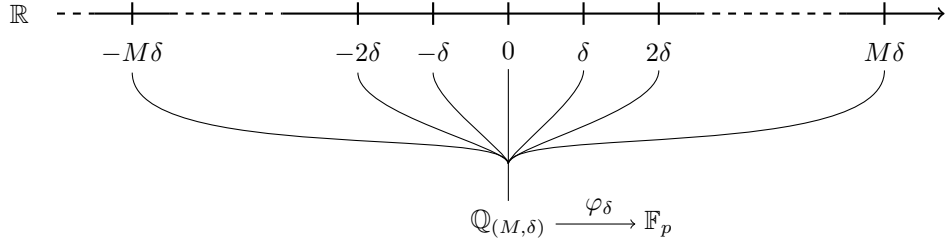

Like~\cite{Catrina2010,Cock2015}, we approach the computation of our public function using fixed-point arithmetic represented using elements of $\mathbb{F}_p$. More precisely, given $M\in\mathbb{N}_+$ we let $\mathbb{Z}_M^\pm=\{x\in\mathbb{Z}\mid -M\leq x\leq M\}$. Then for a scaling factor $\delta>0$, we let
\begin{equation*}
  \mathbb{Q}_{(M,\delta)}=\{ x\delta \mid x\in\mathbb{Z}_M^\pm\},
\end{equation*}
which is also illustrated in Figure~\ref{fig:fixedPoint}.
There is a natural bijection between $\mathbb{Q}_{(M,\delta)}$ and $\mathbb{Z}_M^\pm$, and in order to map fixed-point numbers in $\mathbb{Q}_{(M,\delta)}$ to elements of $\mathbb{F}_p$, we use the map $\varphi'\colon\mathbb{Z}_M^\pm\rightarrow\mathbb{F}_p$ given by $\varphi'(x)=x\bmod{p}$. This is injective as long as $M<p-M$; that is, as long as $p>2M$. In this case, the map $\psi\colon\mathbb{F}_p\rightarrow\mathbb{Z}_M^\pm\cup\{\bot\}$ defined by
\begin{equation*}
  \psi(x)=
  \begin{cases}
    x & x\in\{0,1,\ldots,M\}\\
    x-p & x\in\{p-M,p-M+1,\ldots,p-1\}\\
    \bot & \text{otherwise}
  \end{cases}.
\end{equation*}
is well-defined, and we have $\psi\circ\varphi'=\mathrm{Id}$.

\begin{remark}\label{rem:maximalM}
  If we let $M=\lfloor\frac{p}{2}\rfloor=\frac{p-1}{2}$ for $p>2$, each element of $\mathbb{F}_p$ corresponds to an element of $\mathbb{Z}_M^\pm$. In this case, we can omit $\bot$ from the definition of $\psi$, meaning that $\psi=(\varphi')^{-1}$.
  To simplify notation, we will assume this case in the remainder of the work.
\end{remark}

Rather than working directly with $\varphi'$ and $\psi$, we define $\varphi_\delta\colon\mathbb{Q}_{(M,\delta)}\rightarrow\mathbb{F}_p$ by $\varphi_\delta(x\delta)=\varphi'(x)$. Our assumption from Remark~\ref{rem:maximalM} further implies that $\varphi_\delta^{-1}(x)=\delta\psi(x)$.
Later on, we will need the properties of $\varphi_\delta$ and appropriate inverses when applied to sums and products, so we record these in the following lemma.

\begin{lemma}\label{lem:phiProperties}
  Let $p>2$ be a prime, and let $M=(p-1)/2$ as in Remark~\ref{rem:maximalM}. If $a_1,a_2,\ldots,a_n\in\mathbb{Q}_{(M,\delta)}\subseteq\mathbb{R}$, the map $\varphi_\delta$ satisfies
  \begin{enumerate}
    \item If $\sum_{i=1}^n a_i\in\mathbb{Q}_{(M,\delta)}$, then $\varphi^{-1}_{\delta}\big(\sum_{i=1}^n \varphi_\delta(a_i)\big)=\sum_{i=1}^n a_i$.
    \item If $\prod_{i=1}^n a_i\in\mathbb{Q}_{(M,\delta^n)}$, then $\varphi^{-1}_{\delta^n}\big(\prod_{i=1}^n \varphi_\delta(a_i)\big)=\prod_{i=1}^n a_i$.
  \end{enumerate}
\end{lemma}
\begin{IEEEproof}
  Write $a_i=\delta x_i$ for some $x_i\in\mathbb{Z}_M^\pm$. We then have
  \begin{align*}
    \sum_{i=1}^n\varphi_\delta(a_i)
    &= \sum_{i=1}^n (x_i\bmod{p})\\
    &= \left(\sum_{i=1}^n x_i\right)\bmod{p}
      = \varphi_\delta\left({\sum_{i=1}^n a_i}\right),
  \end{align*}
  where the last equality follows from the assumption $\sum_{i=1}^n a_i\in\mathbb{Q}_{(M,\delta)}$. The proof for products is similar.
\end{IEEEproof}

\subsection{Model assumptions}
Throughout, we will assume that the participants can be passively corrupted -- sometimes referred to as \emph{`honest, but curious'}. This means that they will follow protocols honestly, but try to use the information seen during the protocol to extract additional information about the inputs of the other participant.

We do, however, not put any restriction on the computational power of the adversary in our formal security analysis. That is, our privacy results will show perfect privacy under the assumption that $\mathcal{F}_\mathrm{BT}$ produces Beaver triples with perfect security.
While the assumption that $\mathcal{F}_\mathrm{BT}$ is perfect might seem strong, it does mean that the protocols themselves do not limit the achievable security level. Instead, the security level in practice will be determined by the security level of the Beaver triple generation.

\section{Semi-private computation of correlation}\label{sec:semi-priv-comp}
We assume that participants $P_1$ and $P_2$ each hold samples $(x_j^{(1)})_{j=1}^n$ and $(x_j^{(2)})_{j=1}^n$ of their respective random variables with $x_j^{(i)}\in\mathbb{R}$. Their goal is to compute the sample (Pearson) correlation coefficient
\begin{equation}\label{eq:correlation}
  r=\frac{1}{n-1}\sum_{j=1}^n\left(\frac{x_j^{(1)}-\bar{x}^{(1)}}{s^{(1)}}\right)\left(\frac{x_j^{(2)}-\bar{x}^{(2)}}{s^{(2)}}\right),
\end{equation}
where $s^{(i)}=\sqrt{\frac{1}{n-1}\sum_{j=1}^n(x_j^{(i)}-\bar{x}^{(i)})^2}$ is the sample standard deviation for the data held by $P_i$, and $\bar{x}^{(i)}$ denotes the sample mean of $(x_j^{(i)})_{j=1}^n$. We use the notation $z_j^{(i)}=(x_j^{(i)}-\bar{x}^{(i)})/s^{(i)}$, and note that $P_i$ can compute $z_j^{(i)}$ locally for every $j\in\{1,2,\ldots,n\}$.
With this notation, \eqref{eq:correlation} reduces to the computation of a scalar product $r=\mathbf{z}^{(1)}\cdot\mathbf{z}^{(2)}/(n-1)$ of privately held vectors.

In order to use techniques from MPC, the participants will round each $z_j^{(i)}$ to the nearest element of $\mathbb{Q}_{(M,\delta)}$ (where appropriate choices of $M$ and $\delta$ are discussed in a later section). That is, $P_i$ finds $\varepsilon_j^{(i)}\in [-\frac{\delta}{2},\frac{\delta}{2}]$ such that
\begin{equation}\label{eq:zWithError}
  z_j^{(i)}=\tilde{z}_j^{(i)}+\varepsilon_j^{(i)}
\end{equation}
with $\tilde{z}_j^{(i)}\in\mathbb{Q}_{(M,\delta)}$.
For convenience, we also define $\tilde{r}$ to be the approximation of the sample correlation when using $\mathbb{Q}_{(M,\delta)}$ to represent the $z_j^{(i)}$. That is, we let
\begin{equation}\label{eq:approxCorrelation}
  \tilde{r}=\frac{1}{n-1}\sum_{j=1}^n \tilde{z}_j^{(1)}\tilde{z}_j^{(2)}.
\end{equation}
Based on these definitions, we can rewrite~\eqref{eq:correlation} as
\begin{align}\label{eq:correlationExpansion}
  r &=\frac{1}{n-1}\left(\sum_{j=1}^n z_j^{(1)}z_j^{(2)}\right)\nonumber\\
    &=\tilde{r}+\frac{1}{n-1}\left(\sum_{i=1}^2\sum_{j=1}^n z_j^{(i)}\varepsilon_j^{(3-i)} + \sum_{j=1}^n \varepsilon_j^{(1)}\varepsilon_j^{(2)}\right)
\end{align}
From this, we see that $r$ splits into the approximated sample correlation $\tilde{r}$ plus terms related to the rounding errors. Thus, this leads to a method of computing $r$ in which the participants do an approximation using fixed-point numbers and then accept the leakage of the remaining sums in~\eqref{eq:correlationExpansion}. Before presenting the details, we prove two lemmata that justify the parameter choices.

\begin{lemma}\label{lem:maxError}
  Let $\delta>0$ be fixed, and let $\tilde{r}$ be as in~\eqref{eq:approxCorrelation}. If $R\in \mathbb{R}$ satisfies $|\tilde{z}_j^{(i)}|\leq R$ for all $i\in\{1,2\}$ and all $j\in\{1,2,\ldots,n\}$, then
  \begin{equation*}
    |r-\tilde{r}|\leq \frac{n\delta}{n-1}\left(R+\frac{\delta}{4}\right).
  \end{equation*}
\end{lemma}

\begin{IEEEproof}
  The bound follows by substituting all $\varepsilon_j^{(i)}$ in~\eqref{eq:correlationExpansion} by $\frac{\delta}{2}$ and all $z_j^{(i)}$ by $R$.
\end{IEEEproof}

\begin{lemma}\label{lem:parameterChoices}
  Let $\delta>0$ be fixed, and let $R$ be as in Lemma~\ref{lem:maxError} with the additional assumption that $R\leq \frac{n}{\delta}$. If $M\geq\frac{n-1}{\delta^2}+n(\frac{R}{\delta}+\frac{1}{4})$, then
  \begin{equation*}
    \tilde{r}=\frac{1}{n-1}\,\varphi_{\delta^2}^{-1}\left(\sum_{i=1}^n\varphi_\delta(\tilde{z}_j^{(1)})\varphi_\delta(\tilde{z}_j^{(2)})\right).
  \end{equation*}
\end{lemma}
\begin{IEEEproof}
  First note that the assumptions imply that $M\geq R^2$. Thus, $\tilde{z}_j^{(1)}\tilde{z}_j^{(2)}\in\mathbb{Q}_{(M,\delta^2)}$ for every $j$. Lemma~\ref{lem:phiProperties} then ensures that $\varphi_\delta(\tilde{z}_j^{(1)})\varphi_\delta(\tilde{z}_j^{(2)})=\varphi_{\delta^2}(\tilde{z}_j^{(1)}\tilde{z}_j^{(2)})$. Now, since the true sample correlation satisfies $|r|\leq 1$, we can combine this with Lemma~\ref{lem:maxError} to obtain
  \begin{equation*}
    (n-1)|\tilde{r}|\leq (n-1)|r|+n\delta\left(R+\frac{\delta}{4}\right)\leq \delta^2M.
  \end{equation*}
  This implies that $(n-1)|\tilde{r}|\in\mathbb{Q}_{(M,\delta^2)}$, and the result follows from the first part of the proof combined with Lemma~\ref{lem:phiProperties} applied to $(\tilde{z}_j^{(1)}\tilde{z}_j^{(2)})_{j=1}^n$.
\end{IEEEproof}

The proposed procedure is summed up in Protocol~\ref{prot:revealErrors} on page~\pageref{prot:revealErrors}.

\begin{protocol}[htb]{Exact correlation with controlled leakage}{
      This protocol allows two participants $P_1$ and $P_2$ to compute their exact sample correlation $r$. Before the protocol, the participants have agreed on parameters $R,M,\delta$ such that $|\tilde{z}_j^{(i)}|\leq R\leq \frac{n}{\delta}$ for all $i,j$, and such that $(n-1)/\delta^2+n(R/\delta+1/4)\leq M=(p-1)/2$ for some prime $p$. The participants are assumed to have access to a functionality $\mathcal{F}_{BT}$ for generating Beaver triples.
  }\label{prot:revealErrors}
  \item\label{item:revealErrors}
  $P_i$ locally computes $s^{(i)}=\sqrt{\frac{1}{n-1}\sum_{j=1}^n(x_j^{(i)}-\bar{x})^2}$ and $z_j^{(i)}=(x_j^{(i)}-\bar{x}^{(i)})/s^{(i)}$. It then approximates each $z_j^{(i)}$ by a fixed-point number $\tilde{z}_j^{(i)}\in\mathbb{Q}_{(M,\delta)}$ like in~\eqref{eq:zWithError}.
  $P_i$ sends $(\varepsilon_j^{(i)})_{j=1}^n$ to $P_{3-i}$.

  \item\label{item:revealErrorSum} From the $\varepsilon_j^{(3-i)}$ received in the previous step, $P_i$ locally computes $\sum_{j=1}^n z_j^{(i)}\varepsilon_j^{(3-i)}$ and reveals the result to $P_{3-i}$.
  
  \item\label{item:createShares}
  For each $j\in\{1,2,\ldots,n\}$, $P_i$ creates and distributes sharings $\share{\varphi_\delta(\tilde{z}_j^{(i)})}$.

  \item\label{item:computeProd}
  The participants compute $\share{a}=\sum_{j=1}^n\share{\varphi_\delta(\tilde{z}_j^{(1)})}\share{\varphi_\delta(\tilde{z}_j^{(2)})}$ using Beaver triples and local addition of shares.

  \item\label{item:revealShares}
  $P_i$ reveals $[a]_i$ to $P_{3-i}$, and both participants recover $a$.

  \item\label{item:ouput}
  Each participant outputs $(\varphi_{\delta^2}^{-1}(a)+\sum_{i=1}^2\sum_{j=1}^nz_j^{(i)}\varepsilon_j^{(3-i)}+\sum_{j=1}^n\varepsilon_j^{(1)}\varepsilon_j^{(2)})/(n-1)$.
\end{protocol}

\subsection{Correctness}\label{sec:correctness}
The correctness of Protocol~\ref{prot:revealErrors} follows by noticing that the assumptions on $M$, $\delta$ and $R$ imply $\varphi_{\delta^2}^{-1}(a)=(n-1)\tilde{r}$ by Lemma~\ref{lem:parameterChoices}. Thus, the value output by the participants is exactly the same as in~\eqref{eq:correlationExpansion}.

\subsection{Privacy}
\begin{figure*}[p]
  \centering
  \includegraphics[scale=0.75]{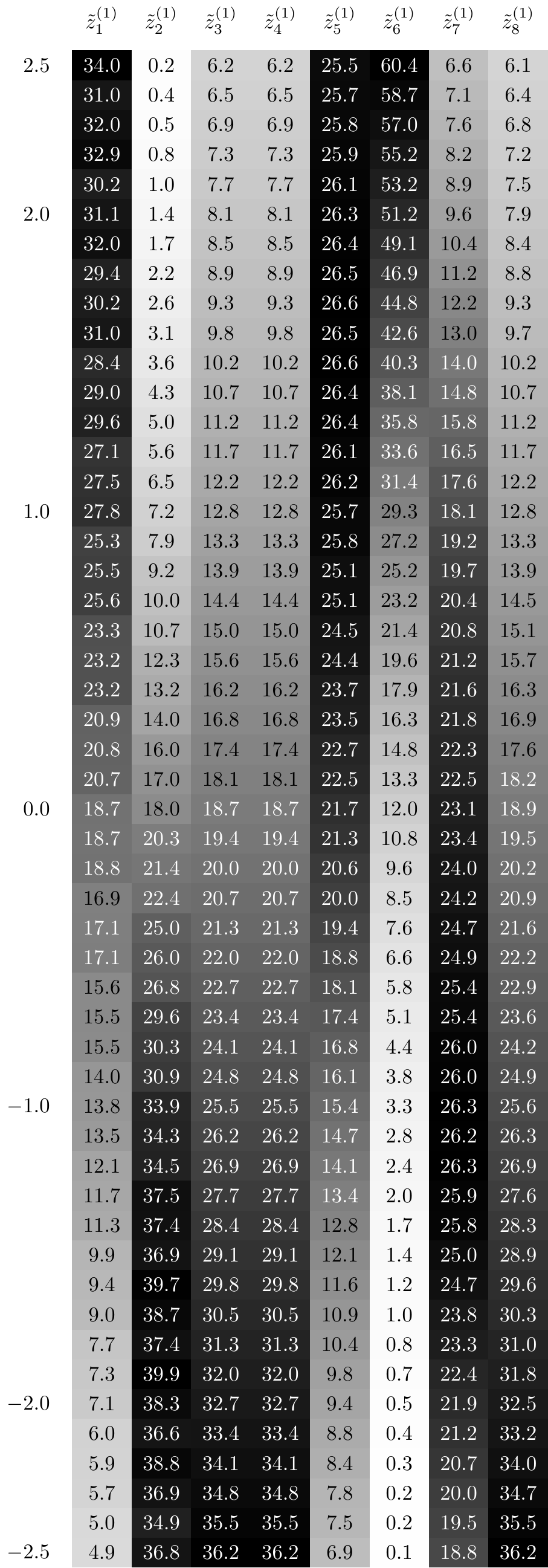}
  \hglue 1cm
  \includegraphics[scale=0.75]{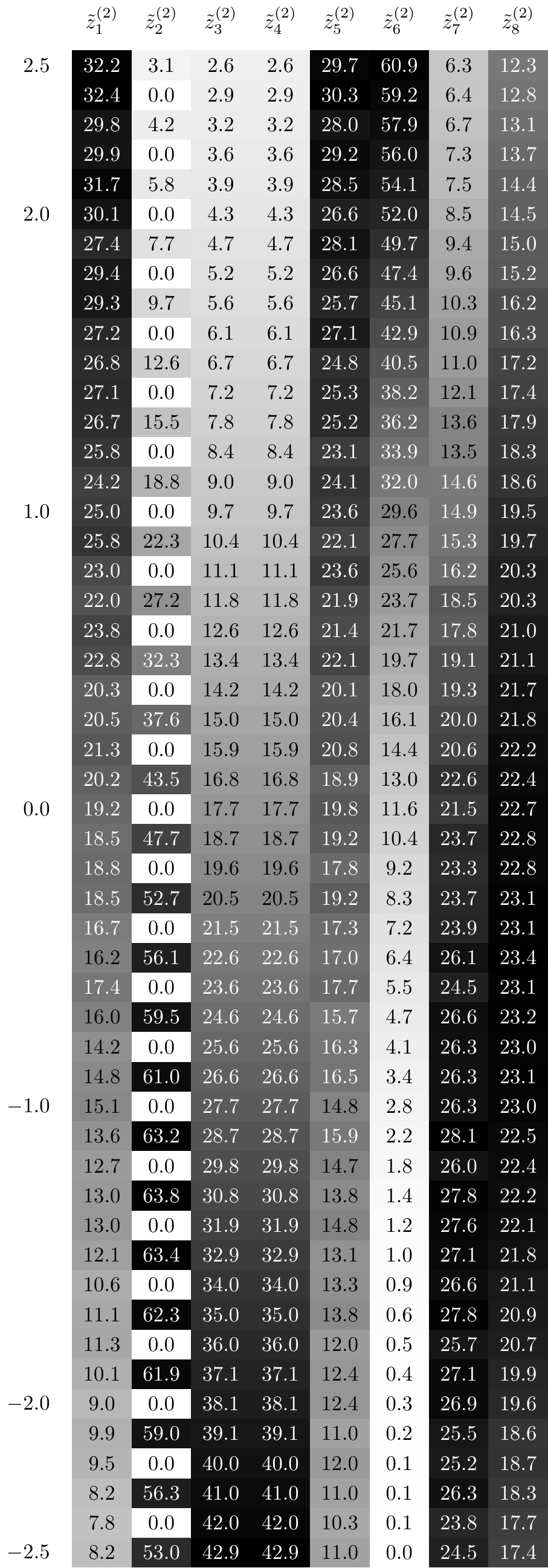}
  \caption{Leaked information in Example~\ref{ex:leakageAnalysis} from the view of $P_2$ (on the left) and $P_1$ (on the right).
      Each column corresponds to one of the variables $\tilde{z}_j^{(i)}$, and each row corresponds to one of the possible elements of $\mathbb{Q}_{(M,\delta)}$ as indicated by the scale on the left. The cell corresponding to $(a,\tilde{z}_j^{(i)})$ gives the relative frequency in \emph{per mille} (‰) of  $\tilde{z}_j^{(i)}=a$ occurring in the valid solutions of Example~\ref{ex:leakageAnalysis}. The cell colour indicates the ratio of the cell value to the maximum in the same column, with $0$ being pure white and the maximum being pure black.}
  \label{fig:digitDistribution}
\end{figure*}

We now prove that Protocol~\ref{prot:revealErrors} is semi-private in the sense of Definition~\ref{defi:normalSecurity}. That is, we consider the real and ideal worlds where the leaker $\mathcal{L}$ and its leakage function $\ell$ is defined based on steps~\ref{item:revealErrors} and \ref{item:revealErrorSum} of the protocol.
\begin{prop}\label{prop:protRevealErrorsSecure}
  Protocol~\ref{prot:revealErrors} implements computation of the Pearson correlation with controlled leakage for leakage function
  \begin{equation*}
    \ell(\mathbf{x}^{(1)},\mathbf{x}^{(2)}) \!=\!
    \bigg(\! (\varepsilon_j^{(1)}, \varepsilon_j^{(2)})_{j=1}^n, \sum_{j=1}^n z_j^{(1)}\varepsilon_j^{(2)}, \sum_{j=1}^n z_j^{(2)}\varepsilon_j^{(1)} \!\bigg).
  \end{equation*}
\end{prop}
\begin{IEEEproof}
  As described in Section~\ref{sec:mpc}, we consider the real world, where participants follow Protocol~\ref{prot:revealErrors} with the help of the functionality $\mathcal{F}_\mathrm{BT}$ for generating Beaver triples and compare this against the ideal world, where computation of $r$ is performed by the ideal functionality $\mathcal{F}_r$, and protocol messages are generated by some simulator $\mathcal{S}$. Recall that this includes a leaker functionality $\mathcal{L}$ in the ideal world, and a null sink $\mathcal{N}$ in the real world.
  We give an explicit description of a simulator $\mathcal{S}$ such that the two settings are perfectly indistinguishable.
  For notational ease, we assume that $P_1$ is passively corrupt. The case where $P_2$ is corrupt follows by symmetry.

  First, $P_1$ sends its input to the functionality that is either $\mathcal{L}$ or $\mathcal{N}$, but since $P_1$ does not receive any response in either the real or ideal worlds, the transcript will be identical in both cases.

  Considering then steps~\ref{item:revealErrors} and \ref{item:revealErrorSum}, $\mathcal{S}$ can use the values of $(\varepsilon_j^{(2)})_{j=1}^n$ and $\sum_{j=1}^n z_j^{(2)}\varepsilon_j^{(1)}$ it received from $\mathcal{L}$, and what is received by $P_1$ is identical in both worlds.

  In step~\ref{item:createShares}, $\mathcal{S}$ chooses $\share{\varphi_\delta(\tilde{z}_j^{(2)})}_1$ uniformly at random in $\mathbb{F}_p$. These are distributed identically to the shares that $P_1$ would receive in the ideal world.
  
  Moving on to step~\ref{item:computeProd}, the Beaver triples $(\share{u},\share{v},\share{w})$ are generated by $\mathcal{S}$, and $P_1$ receives the corresponding shares. Since $\varphi_\delta(\tilde{z}_j^{(1)})-u$ is revealed during computation of the product $\share{\varphi_\delta(\tilde{z}_j^{(1)})}\share{\varphi_\delta(\tilde{z}_j^{(2)})}$ and $u$ is known to $\mathcal{S}$, it can extract $\varphi_\delta(\tilde{z}_j^{(1)})$ and thereby compute $z_j^{(1)}$ for each $j\in\{1,2,\ldots,n\}$. In addition, note that this information also allows $\mathcal{S}$ to compute $\share{a}_1$.
  It passes $(z_j^{(1)})_{j=1}^n$ as inputs to $\mathcal{F}_r$ and receives the output $r$.

  In step~\ref{item:revealShares}, $\mathcal{S}$ sets
  \begin{equation*}
    \share{a}_2=\varphi_{\delta^2}\bigg((n-1)r-\sum_{i=1}^2\sum_{j=1}^n z_j^{(i)}\varepsilon_j^{(3-i)}-\sum_{j=1}^n\varepsilon_j^{(1)}\varepsilon_j^{(2)}\bigg)-\share{a}_1
  \end{equation*}
  and sends this value to $P_1$. By~\eqref{eq:correlationExpansion}, this implies $\share{a}_1+\share{a}_2=(n-1)\tilde{r}$ like in the real world, and furthermore, there is only one choice of $\share{a}_2$ given $\tilde{r}$ and $\share{a}_1$.

  In summary, the messages produced by $\mathcal{S}$ are perfectly indistinguishable from those seen in a real execution of Protocol~\ref{prot:revealErrors}, proving the result.
\end{IEEEproof}
To illustrate the use of Protocol~\ref{prot:revealErrors}, we give two examples. The first focuses on the procedure itself and the second illustrates the kind of information that is leaked during the protocol.
Note that the scaling factor used here is extremely small in order to make the examples easier to follow. We treat appropriate choices of parameters in Section~\ref{sec:bounds-pract-impl}.
\begin{example}\label{ex:protocolExecution}
  Assume that $P_1$ and $P_2$ hold samples
  \begin{align*}
    \mathbf{x}^{(1)} &=\resizebox{0.85\linewidth}{!}{(2.113, -0.906, -0.546, -1.550, 1.770, 4.002, -0.135, \phantom{-}0.606)}\\
    \mathbf{x}^{(2)} &=\resizebox{0.85\linewidth}{!}{(0.647, -2.108, -0.479, -1.751, 0.442, 2.105, -0.836, -1.748)}
  \end{align*}
  respectively. Computing their $z$-scores in Step~\ref{item:revealErrors} of Protocol~\ref{prot:revealErrors}, they obtain (if rounded to three decimal places)
  \begin{align*}
    \mathbf{z}^{(1)} &= \resizebox{0.85\linewidth}{!}{(0.781, -0.852, -0.657, -1.200, 0.595, 1.802, -0.435, -0.034)}\\
    \mathbf{z}^{(2)} &= \resizebox{0.85\linewidth}{!}{(0.765, -1.129, -0.009, -0.884, 0.624, 1.768, -0.254, -0.882)}
  \end{align*}
  Using scaling factor $\delta=0.1$ and setting $R=2.5$, we must choose $M\geq 902$ to satisfy the assumptions of the protocol. The smallest choice of prime is then $p=1811$. As described in Remark~\ref{rem:maximalM}, this implies $M=905$.

  With these parameters in place, $P_1$ rounds the entries of $\mathbf{z}^{(1)}$ to the nearest elements in $\mathbb{Q}_{(M,\delta)}$, obtaining
  \begin{equation*}
    \tilde{\mathbf{z}}^{(1)}= (0.8, -0.9, -0.7, -1.2, 0.6, 1.8, -0.4, 0.0).
  \end{equation*}
  When this is mapped to $\mathbb{F}_{1811}$, it yields $\varphi_{\delta}(\tilde{\mathbf{z}}^{(1)})=(8,1802,1804,1799,6,18,1807,0)\in\mathbb{F}_{1811}^8$. Similarly, $P_2$ ends up computing $\varphi_{\delta}(\tilde{\mathbf{z}}^{(2)})=(8,1800,0,1802,6,18,1808,1802)\in\mathbb{F}_{1811}^8$. While performing these computations, they have also recorded and revealed the error vectors $\boldsymbol\varepsilon^{(i)}=\mathbf{z}^{(i)}-\tilde{\mathbf{z}}^{(i)}$ as well as the values $\sum_{j=1}^nz_j^{(1)}\varepsilon_j^{(2)}\approx -0.079$ and $\sum_{j=1}^nz_j^{(2)}\varepsilon_j^{(1)}\approx -0.029$.

  They then create sharings of these, and use the Beaver triples to compute shares of the entrywise product $\mathbf{y} =\varphi_\delta(\tilde{\mathbf{z}}^{(1)})\odot\varphi_\delta(\tilde{\mathbf{z}}^{(2)})$, giving
  \begin{equation*}
    \mathbf{y}= (64, 99, 0, 108, 36, 324, 12, 0)\in\mathbb{F}_{1811}^8.
  \end{equation*}
  Now, $P_i$ locally computes the sum $\sum_{j=1}^n\share{y_j}_i$, and reveals this in step~\ref{item:revealShares}. Thus, both participants recover $\sum_{j=1}^n y_j= 643$. Mapping this back to $\mathbb{Q}_{(M,\delta)}$, they get $\varphi_{\delta^2}^{-1}(643)=6.43$. Since the error vectors have previously been revealed, the participants are able to locally compute the sum $\sum_{j=1}^n\varepsilon_j^{(1)}\varepsilon_j^{(2)}\approx 0.007$. Combining all of this, they recover
  \begin{align*}
    \frac{1}{n-1}\bigg(&\varphi_{\delta^2}^{-1}(643)+\sum_{i=1}^2\sum_{j=1}^nz_j^{(i)}\varepsilon_j^{(3-i)}+\sum_{j=1}^n\varepsilon_j^{(1)}\varepsilon_j^{(2)}\bigg)\\
                      &\approx \frac{1}{7}(6.43-0.079-0.029+0.007) \approx 0.904,
  \end{align*}
  which is equal to the exact sample correlation apart from errors introduced by the rounding used in this exposition.
\end{example}
\begin{example}\label{ex:leakageAnalysis}
  We consider the same protocol execution as in Example~\ref{ex:protocolExecution}, and analyse it from the view of $P_2$. During the protocol, $P_2$ learns $(\varepsilon_j^{(1)})_{j=1}^n$, $\sum_{j=1}^n z_j^{(1)}z_j^{(2)}$, and $\sum_{j=1}^n z_j^{(1)}\varepsilon_j^{(2)}$.
  Observe that learning the latter two sums correspond to two linear equations in the $n=8$ unknowns $z_1^{(1)},z_2^{(1)},\ldots,z_8^{(1)}$.
  Rather than working with $z_j^{(1)}$ as unknowns, we will use $(\varepsilon_j^{(1)})_{j=1}^n$ to rewrite the system to an equivalent system in the unknowns $\tilde{z}_j^{(1)}$. More precisely, $P_2$ can form the system
  \begin{equation}\label{eq:p2EqSystem}
    \left\{
      \begin{array}{@{}r@{\,}l}
        \sum_{j=1}^n \tilde{z}_j^{(1)} z_j^{(2)} &= \sum_{j=1}^n z_j^{(1)}z_j^{(2)}-\sum_{j=1}^n \varepsilon_j^{(1)}z_j^{(2)}\\
        \sum_{j=1}^n \tilde{z}_j^{(1)} \varepsilon_j^{(2)} &= \sum_{j=1}^n z_j^{(1)}\varepsilon_j^{(2)}-\sum_{j=1}^n \varepsilon_j^{(1)}\varepsilon_j^{(2)} 
      \end{array}
    \right.
  \end{equation}
  where we note that the right-hand side is known to $P_2$. Hence, from the point of view of $P_2$, \eqref{eq:p2EqSystem} is a linear system of two equations in the $n=8$ unknowns $\tilde{z}_1^{(1)},\tilde{z}_2^{(1)},\ldots,\tilde{z}_8^{(1)}$. Solving this with the data samples from Example~\ref{ex:protocolExecution} yields six free variables $(\tilde{z}_j^{(1)})_{j=3}^8$, but not all assignments of values to these give a solution that is consistent with $\tilde{z}_j^{(1)}\in\mathbb{Q}_{(M,\delta)}$. In addition, $P_2$ will know the parameter $R=2.5$ used in the protocol, meaning additionally that $\lvert\tilde{z}_j^{(1)}\rvert\leq 2.5$. Since this example is relatively small, we can go through all possible assignments of $(\tilde{z}_j^{(1)})_{j=3}^8$ satisfying these conditions and record those that also imply $\tilde{z}_1^{(1)},\tilde{z}_2^{(1)}\in \mathbb{Q}_{(M,\delta)}\cap [-2.5,2.5]$.
  There are a total of $3624355$ such solutions, when accepting a small additive error ($10^{-12}$) in each variable to allow for precision loss in the floating-point computations. Figure~\ref{fig:digitDistribution} on page~\pageref{fig:digitDistribution} shows the frequencies of values for each variable in these solutions as well as the corresponding analysis from the view of $P_1$. This illustrates, that the participants cannot determine the actual data held by the other participant, but may gain some insights about the distribution of the individual sample points.
  In principle, one could perform similar frequency analysis for pairs, triples etc. of variables in a hope to gain further information, but we shall refrain from doing so in this work.
\end{example}

Whether the leakage from Protocol~\ref{prot:revealErrors} is acceptable or not depends on the use case as well as the distributions of the participants. As the example below shows, it may lead to a complete leakage in certain cases.
\begin{example}
  Consider the case where the samples of $P_2$ come from a discrete distribution whose probability mass function has exactly two points in its support. That is, $x_j^{(2)}\in\{a,b\}$ for some real numbers $a\neq b$. In this example, we assume that $P_1$ somehow knows that the distribution of $P_2$ has this form (but without knowing the values of $a$ and $b$).

  Writing $a=\tilde{a}+\varepsilon_a$ for $\tilde{a}\in\mathbb{Q}_{(M,\delta)}$ like in~\eqref{eq:zWithError}, and denoting by $\mathcal{J}_a$ the indices $j$ such that $x_j^{(2)}=a$, we have that $\varepsilon_j^{(2)}=\varepsilon_a$ for all $j\in\mathcal{J}_a$. With similar notation for $b$, we get $\varepsilon_j^{(2)}=\varepsilon_b$ for all $j\in\mathcal{J}_b$.
  Thus, unless $\varepsilon_a=\varepsilon_b$, $P_1$ can determine $\mathcal{J}_a$ and $\mathcal{J}_b$ from $(\varepsilon_j^{(2)})_{j=1}^n$.
  During Protocol~\ref{prot:revealErrors}, $P_1$ learns the equation $\sum_{j=1}^nz_j^{(1)}z_j^{(2)}=r$ and the value of $\sum_{j=1}^n z_j^{(1)}\varepsilon_j^{(2)}$. Combining this with the knowledge of $\mathcal{J}_a$ and $\mathcal{J}_b$ leads to the system
  \begin{equation*}
    \left\{
      \begin{array}{@{}l}
        a\left(\sum_{j\in\mathcal{J}_a}z_j^{(1)}\right) + b\left(\sum_{j\in\mathcal{J}_b}z_j^{(1)}\right) \!= r\\
        a\left(\sum_{j\in\mathcal{J}_a}\varepsilon_j^{(1)}\right) + b\left(\sum_{j\in\mathcal{J}_b}\varepsilon_j^{(1)}\right) \!= \sum_{j=1}^n z_j^{(1)}\varepsilon_j^{(2)},\\
      \end{array}
    \right.
  \end{equation*}
  which from the point of view of $P_1$ is a system of two linear equations in two unknowns, $a$ and $b$. Unless these equations are linearly dependent, $P_1$ can determine $a$ and $b$ uniquely, and thus fully learns the data of $P_2$. Note, however, that this attack relies on $P_1$ knowing that the samples of $P_2$ have this special form.
\end{example}

\subsection{Complexity}
We now turn our attention to the cost of Protocol~\ref{prot:revealErrors} in terms of computation and communication. For this analysis, we consider one addition or multiplication in $\mathbb{R}$ or $\mathbb{F}_p$ to be a single operation, and we will count $\mathbb{R}$-operations and $\mathbb{F}_p$-operations separately. Similarly, we count the transmission of symbols from $\mathbb{R}$ and $\mathbb{F}_p$ separately\footnote{In practice, the elements of $\mathbb{R}$ will likely be represented by floating-point numbers, but we ignore this detail here.}.
We do not include the cost of generating the Beaver triples in this analysis since that will depend on the specific implementation of $\mathcal{F}_{BT}$. We note, however, that as a specific example~\cite{Boyle20} provides a way to generate $n$ triples using $\mathcal{O}(c^2n\log n)$ $\mathbb{F}$-operations, where $c$ is a small integer, as well as a number of calls to a pseudorandom generator. For the details, see~\cite{Boyle20}.
\begin{prop}\label{prop:complexity}
  During Protocol~\ref{prot:revealErrors}, the participants consume $n$ Beaver triples. Furthermore, each participant
  \begin{itemize}
    \item performs $\mathcal{O}(n)$ $\mathbb{R}$-operations and $\mathcal{O}(n)$ $\mathbb{F}_p$-operations
    \item transmits $\mathcal{O}(n)$ elements of $\mathbb{R}$ and $\mathcal{O}(n)$ elements of $\mathbb{F}_p$
  \end{itemize}
\end{prop}
\begin{IEEEproof}
  In step~\ref{item:revealErrors}, the computation of $s^{(i)}$ and $z_j^{(i)}$ takes a total of $6n+\mathcal{O}(1)$ operations in $\mathbb{R}$ for each participant, where the constant term covers the cost of computing the square root in $s^{(i)}$. In addition, each participant transmits the $n$ error values. In the next step, computing the sum requires $2n-1$ $\mathbb{R}$-operations, and revealing it requires transmission of one element of $\mathbb{R}$.
  Then in step~\ref{item:createShares}, creating each of the $n$ sharings require $\mathcal{O}(1)$ $\mathbb{F}_p$-operations, and distributing it requires transmission of a single $\mathbb{F}_p$-element.
  Moving to step~\ref{item:computeProd}, each product computed using Beaver triples costs at most $8$ operations in $\mathbb{F}_p$ per participant as well as transmission of $2$ $\mathbb{F}_p$-elements. Thus, the computation of $a$ gives a total of at most $9n-1$ $\mathbb{F}_p$-operations and transmission of $2n$ $\mathbb{F}_p$-elements.
  When revealing $\share{a}_i$ afterwards, this is the transmission of a single element of $\mathbb{F}_p$.
  Finally, for the computation of the output, observe that the participants already know the values of $\sum_{j=1}^nz_j^{(i)}\varepsilon_j^{(3-j)}$. Furthermore, the computational cost of computing $\varphi_\delta^{-1}(a)$ is independent of $n$, meaning that the participants can compute the result using $2n+\mathcal{O}(1)$ operations.

  Summing all of these yields $10n+\mathcal{O}(1)$ $\mathbb{R}$-operations, $\mathcal{O}(n)$ $\mathbb{F}_p$-operations as well as transmission of $n+1$ and $3n+1$ elements of $\mathbb{R}$ and $\mathbb{F}_p$, respectively.
\end{IEEEproof}

\subsection{Quantifying the leakage}
If the computation of the sample correlation is performed by a trusted third party and then revealed to the participants, the information revealed to $P_1$ about the input of $P_2$ is
\begin{equation*}
  \lambda = I\big(\mathbf{X}^{(2)}; \mathbf{X}^{(1)}, r(\mathbf{X}^{(1)},\mathbf{X}^{(2)})\big)
\end{equation*}
where $I$ denotes the mutual information between random variables.
If, on the other hand, the participants follow Protocol~\ref{prot:revealErrors}, the information gained by $P_1$ is
\begin{align*}
  \lambda' = I(\mathbf{X}^{(2)}; \mathbf{X}^{(1)}, \tilde{r}(\mathbf{X}^{(1)},\mathbf{X}^{(2)}),& \mathbf{E}^{(1)}, \mathbf{E}^{(2)},\\
  &\mathbf{Z}^{(1)}\cdot\mathbf{E}^{(2)}, \mathbf{Z}^{(2)}\cdot\mathbf{E}^{(1)}),
\end{align*}
where $\mathbf{E}^{(i)}$ is the random variable corresponding to $(\varepsilon_j^{(i)})_{j=1}^n$ and similarly $\mathbf{Z}^{(i)}$ to $(z_j^{(i)})_{j=1}^n$.

This means that in order to fully quantify the additional leakage in Protocol~\ref{prot:revealErrors} compared to a non-leaky protocol in terms of number of bits leaked, one would have to determine (or estimate) $\lambda'-\lambda$. Doing so is highly non-trivial, however. First, the mutual information depends on the distribution of the individual variables. Second, even if specific distributions are assumed for the inputs $\mathbf{X}_1$ and $\mathbf{X}_2$, determining the distributions of the remaining variables is difficult.
For instance, the distribution of $r(\mathbf{X}^{(1)},\mathbf{X}^{(2)})$ appearing in $\lambda$ is known when both $\mathbf{X}^{(i)}$ are one-dimensional and normal~\cite{Hotelling53} -- more precisely when $(X^{(1)}, X^{(2)})$ is bivariate normal -- but its density involves both the gamma function and a hypergeometric series. In our setting, though, it only really makes sense to consider inputs consisting of several samples.

One might expect that obtaining a numerical estimate for $\lambda'-\lambda$ would be much simpler since it is easy to sample all the variables involved as long as the distributions of $\mathbf{X}_1$ and $\mathbf{X}_2$ are known (and one has an efficient way to sample from said distributions). In order to estimate the mutual information from such samples, however, one would first have to estimate the densities of multivariate random variables, and then do a high-dimensional integral involving these densities in order to compute the mutual information.
We consider this to be outside the scope of the current work.

\section{Approximation of the correlation}\label{sec:appr-corr}
This section treats the case where the participants simply approximate their inputs as fixed-point numbers and performs MPC `as usual' -- that is, without ensuring functional privacy as in~\cite{Feigenbaum2006}. This essentially Protocol~\ref{prot:revealErrors}, but without the explicit disclosure of the error values. The resulting procedure is given in Protocol~\ref{prot:approxCorrelation} on page~\pageref{prot:approxCorrelation}. Now, the result is no longer exact, but instead an approximation with an additive error as described by Lemma~\ref{lem:maxError}. As we show in Proposition~\ref{prop:securityApprox}, this approximation reveals no more than the error terms revealed in Protocol~\ref{prot:revealErrors}. In other words, Protocol~\ref{prot:approxCorrelation} is at least as secure as Protocol~\ref{prot:revealErrors}.
The correctness of Protocol~\ref{prot:approxCorrelation} follows by Lemma~\ref{lem:parameterChoices} using similar arguments as in Section~\ref{sec:correctness}.

\begin{protocol}[t]{Approximate correlation with controlled leakage}{
      This protocol allows two participants $P_1$ and $P_2$ to compute an approximation of their sample correlation $r$. Before the protocol, the participants have agreed on parameters $R,M,\delta$ such that $|\tilde{z}_j^{(i)}|\leq R\leq \frac{n}{\delta}$ for all $i,j$, and such that $(n-1)/\delta^2+n(R/\delta+1/4)\leq M=(p-1)/2$ for some prime $p$. 
  }\label{prot:approxCorrelation}
  \item $P_i$ locally computes $s^{(i)}=\sqrt{\frac{1}{n-1}\sum_{j=1}^n(x_j^{(i)}-\bar{x})^2}$ and $z_j^{(i)}=(x_j^{(i)}-\bar{x}^{(i)})/s^{(i)}$.
  
  \item\label{item:shareApprox}
  For each $j\in\{1,2,\ldots,n\}$, $P_i$ creates and distributes sharings $\share{\varphi_\delta(\tilde{z}_j^{(i)})}$.

  \item\label{item:computeApprox}
  The participants compute $\share{a}=\sum_{j=1}^n\share{\varphi_\delta(\tilde{z}_j^{(1)})}\share{\varphi_\delta(\tilde{z}_j^{(2)})}$ using Beaver triples and local addition of shares.

  \item\label{item:outputApprox}
  $P_i$ reveals its share $\share{a}_i$ to $P_{3-i}$, and each participant reconstructs and outputs $\varphi_{\delta^2}^{-1}(a)/(n-1)$.
\end{protocol}

\begin{prop}\label{prop:securityApprox}
  Let $R,M,\delta$ be fixed. Protocol~\ref{prot:approxCorrelation} reveals no more than Protocol~\ref{prot:revealErrors} when the same parameters are used.
\end{prop}
\begin{IEEEproof}
  We consider a game, where a distinguisher inputs $(x_j^{(i)})_{j=1}^n$ for $i=1,2$ and receives a protocol transcript from the view of $P_i$ for some choice of $i\in\{1,2\}$. The transcript is either from an execution of Protocol~\ref{prot:approxCorrelation} or generated by some simulator $\mathcal{S}$ that is given a transcript of Protocol~\ref{prot:revealErrors} from the view of $P_i$ (with the same inputs). We give a simulator $\mathcal{S}$ such that the two cases are perfectly indistinguishable.
  
  First, notice that the shares revealed in steps~\ref{item:shareApprox} and \ref{item:computeApprox} of Protocol~\ref{prot:approxCorrelation} are perfectly indistinguishable from uniformly sampled elements of $\mathbb{F}_p$ by the security of the secret sharing scheme.
  In addition, observe that during the execution of Protocol~\ref{prot:revealErrors}, $P_i$ learns $a=\sum_{j=1}^n\varphi_\delta(\tilde{z}_j^{(1)})\varphi_\delta(\tilde{z}_j^{(2)})$, which is identical to the $a$ recovered in step~\ref{item:outputApprox} of Protocol~\ref{prot:approxCorrelation}. Thus, by creating a random sharing of $a$, $\mathcal{S}$ can generate something that has the same distribution as the shares $\share{a}_1$ and $\share{a}_2$ revealed in step~\ref{item:outputApprox} of Protocol~\ref{prot:approxCorrelation}.
\end{IEEEproof}

\begin{prop}
  During Protocol~\ref{prot:approxCorrelation}, the participants consume $n$ Beaver triples. Furthermore, each participant
  \begin{itemize}
    \item performs $\mathcal{O}(n)$ $\mathbb{R}$-operations and $\mathcal{O}(n)$ $\mathbb{F}_p$-operations
    \item transmits $\mathcal{O}(n)$ elements of $\mathbb{F}_p$ (and no elements of $\mathbb{R}$)
  \end{itemize}
\end{prop}
We omit the proof since it is essentially the same considerations in the proof of Proposition~\ref{prop:complexity}.

\begin{example}
  If the participants in Example~\ref{ex:protocolExecution} had instead used Protocol~\ref{prot:approxCorrelation}, they would not learn the information regarding the error values of the other participant. Thus, they would instead output $\varphi_{\delta^2}^{-1}(643)/7\approx 0.92$ rather than the exact sample correlation.
  Since the scaling factor is relatively small ($\delta=0.1$), the maximal error is relatively large at approximately $0.28$ according to Lemma~\ref{lem:maxError}. Thus, the participants can only conclude that the sample correlation is in the interval $[0.64,1]$.
\end{example}

\subsection{Bounds in practical implementations}\label{sec:bounds-pract-impl}
By choosing $p=4294967291$, calculations in $\mathbb{F}_p$ can be implemented using standard 64-bit unsigned integers, and $p$ is the largest such prime. By rewriting the bound on $M$ from Protocols~\ref{prot:revealErrors} and \ref{prot:approxCorrelation}, we obtain
\begin{equation}\label{eq:boundDelta}
  0\leq \delta^2\left(M-\frac{n}{4}\right)-\delta nR-n+1.
\end{equation}
Thus, by setting $M=\frac{p-1}{2}$ and picking specific choices for $n$ and $R$, we can find the smallest value $\delta_{\min}$ satisfying both $\delta_{\min}\geq 0$ and~\eqref{eq:boundDelta}. Doing so yields
\begin{equation*}
  \delta_{\min}=\frac{-nR+\sqrt{n^2R^2+4(M-n/4)(n-1)}}{2(M-n/4)},
\end{equation*}
and in Table~\ref{tab:deltaBounds} we list values of $\delta_{\min}$ for various values of $n$ and $R$ along with the maximal error as given in Lemma~\ref{lem:maxError}.

\begin{table*}[ht]
  \centering
  \begin{tabular}{*{7}{>{$}r<{$}}}
    \toprule
    & \multicolumn{2}{c}{$R=2.5$} & \multicolumn{2}{c}{$R=5$} & \multicolumn{2}{c}{$R=10$}\\
    \cmidrule(rl){2-3}\cmidrule(rl){4-5}\cmidrule(rl){6-7}
    n & \delta_{\min} & \mathrm{err}_{\max} & \delta_{\min} & \mathrm{err}_{\max} & \delta_{\min} & \mathrm{err}_{\max}\\
    \midrule
    100     & 2.15\cdot 10^{-4} & 5.42\cdot 10^{-4} & 2.15\cdot 10^{-4} & 1.08\cdot 10^{-3} & 2.14\cdot 10^{-4} & 2.17\cdot 10^{-3} \\
    1000    & 6.81\cdot 10^{-4} & 1.71\cdot 10^{-3} & 6.81\cdot 10^{-4} & 3.41\cdot 10^{-3}	& 6.80\cdot 10^{-4} & 6.80\cdot 10^{-3} \\
    10000   & 2.15\cdot 10^{-3} & 5.38\cdot 10^{-3} & 2.15\cdot 10^{-3} & 1.07\cdot 10^{-2}	& 2.13\cdot 10^{-3} & 2.13\cdot 10^{-2} \\
    100000  & 6.77\cdot 10^{-3} & 1.69\cdot 10^{-2} & 6.71\cdot 10^{-3} & 3.36\cdot 10^{-2}	& 6.60\cdot 10^{-3} & 6.60\cdot 10^{-2} \\
    1000000 & 2.10\cdot 10^{-2} & 5.26\cdot 10^{-2} & 2.04\cdot 10^{-2} & 1.02\cdot 10^{-1}	& 1.94\cdot 10^{-2} & 1.94\cdot 10^{-1}\\
    \bottomrule
  \end{tabular}
  \smallskip
  \caption{Minimal scaling factor and maximal error for various choices of $n$ and $R$, when $p=4294967291$.}
  \label{tab:deltaBounds}
\end{table*}

\section{Privacy/precision trade-off}\label{sec:tradeOff}

The information-theoretic privacy guarantee offered by additive secret sharing over finite fields stems from the possibility of choosing values uniformly at random in the field. The unknown shares then act as a one-time pad, and regardless of the shared value each element of the field is equally likely as the share.
Hence, the share reveals nothing about the value of the secret.

Using real numbers, the situation is different. First of all, assume that the secret is in a restricted interval, $s\in[-R,R]$, and that we use standard additive secret sharing. In that case, the shares may in themselves leak information about the secret value. For instance, if $P_1$ receives the share $x_1>0$, then $P_1$ can reduce the possible values of $s$ to the interval $[x_1-R,R]$. This is especially problematic if $x_1$ is very close to $R$.
Even when using more general real secret sharing schemes for two participants, some leakage seems unavoidable. Namely, if $P_1$ can guess the share of $P_2$, this will reveal the secret. In other words, the share of $P_2$ should contain as much entropy as possible, i.e. ideally it should follow the maximum entropy distribution. For finite intervals as above, this is the uniform distribution~\cite[p.\,412]{CoverThomas} as in the finite field case, but one encounters problems like the one above in real number settings.
If, on the other hand, $s\in (-\infty, \infty)$ then the maximum entropy distribution (conditioned on a fixed mean and a finite variance) is the normal distribution~\cite[p.\.413]{CoverThomas}. Since certain values are more likely in a normal distribution, this will imply that certain secrets are more likely -- so some information is leaked to $P_1$.

The discussion above indicates that leakage is likely to be unavoidable in the real number setting, so we compare the two protocols presented in this work with that in mind.
In Protocol~\ref{prot:revealErrors}, the participants reveal certain error terms related to representing their data using fixed-point numbers rather than reals. On the other hand, in Protocol~\ref{prot:approxCorrelation}, they simply use the fixed-point representation, leading to an error in the resulting correlation as described in Lemma~\ref{lem:maxError}. While such an approximation could in principle reveal something about the inputs -- see e.g~\cite{Feigenbaum2006} -- Proposition~\ref{prop:securityApprox} indicates that the leakage is not worse than Protocol~\ref{prot:revealErrors} -- although that is likely a pessimistic bound. Doing so allows them to compute the sample correlation exactly.

These observations show that the participants can trade some of the privacy of their data in order to obtain better precision. Of course, the participants could also obtain the exact value by doing computations in the open (e.g. by simply send their inputs to each other), but Protocol~\ref{prot:revealErrors} provides the same precision with more limited leakage.

\section{Open problems}\label{sec:open-problems}
In the more general case where each participant hold samples from a multivariate random variable -- i.e. the $j$'th sample of $P_i$ has the form $\mathbf{x}_j^{(i)}\in\mathbb{R}^{n_i}$ for some $n_1,n_2$ -- the correlation between them can be represented in a matrix
\begin{equation*}
  M=\left[
    \begin{array}{cc}
      M_1 & M_{12}\\
      M_{12}^\top & M_2
    \end{array}
  \right]
\end{equation*}
where $M_1$ and $M_2$ depend only on the data held by $P_1$ and $P_2$, respectively, and $M_{12}$ contains the `mixed' correlations. Hence, rather than computing a single correlation as in~\eqref{eq:correlation}, the goal is to compute $M_{12}$. Of course, Protocols~\ref{prot:revealErrors} and \ref{prot:approxCorrelation} can be used to compute each entry of this matrix.
Before doing so, however, a more detailed of the information leakage is required, as each variable is part of the computation in several entries. For example, focusing on a single variable $(\mathbf{x}_j^{(1)})_k$ held by $P_1$, computing the correlation with each of the $n_2$ variables of $P_2$ will reveal another two linear equations describing $(\mathbf{x}_j^{(1)})_k$. Although these equations might not be linearly independent, they still represent a more severe information leakage than in the univariate case.
We refrain from making such an analysis in the current work.

In addition to multidimensional data, another direction is the computation of other similarity measures, such as the statistical distance, e.g. the Kullback-Leibler (KL) divergence. This will also require computations on data points where each component is held by different participants. So this problem exhibits a similar structure, but will require a separate analysis of the necessary MPC techniques and the resulting error bounds.

\section{Conclusion}\label{sec:conclusion}
In this work, we presented two methods for computing sample correlation. While not secure in the formal sense -- as in functional privacy -- the information that is leaked can be described precisely in the sense that the participants know exactly which variables they allow leakage of.
We also highlight how the leaky protocol provides improved precision, thus presenting a trade-off between privacy and precision.
In certain practical problems, the security guarantees provided in the current work might be sufficient.
In addition, we demonstrated that even with practical field sizes, the approximation error is not too large.

\bibliographystyle{ieeetr}
\bibliography{ref}

\begin{thebibliography}{10}

\bibitem{liang2018survey}
F.~Liang, W.~Yu, D.~An, Q.~Yang, X.~Fu, and W.~Zhao, ``A survey on big data
  market: Pricing, trading and protection,'' {\em {IEEE} Access}, vol.~6,
  pp.~15132--15154, 2018.

\bibitem{acemoglu2019too}
D.~Acemoglu, A.~Makhdoumi, A.~Malekian, and A.~Ozdaglar, ``Too much data:
  Prices and inefficiencies in data markets,'' tech. rep., National Bureau of
  Economic Research, 2019.

\bibitem{agarwal2019marketplace}
A.~Agarwal, M.~Dahleh, and T.~Sarkar, ``A marketplace for data: An algorithmic
  solution,'' in {\em Proceedings of the 2019 ACM Conference on Economics and
  Computation}, pp.~701--726, 2019.

\bibitem{pinson2022regression}
P.~Pinson, L.~Han, and J.~Kazempour, ``Regression markets and application to
  energy forecasting,'' 2022.

\bibitem{ghorbani2019data}
A.~Ghorbani and J.~Zou, ``Data shapley: Equitable valuation of data for machine
  learning,'' in {\em International Conference on Machine Learning},
  pp.~2242--2251, PMLR, 2019.

\bibitem{pandey2022participation}
S.~R. Pandey, P.~Pinson, and P.~Popovski, ``Participation and data valuation in
  {IoT} data markets through distributed coalitions,'' {\em arXiv preprint
  arXiv:2206.07785}, 2022.

\bibitem{fudenberg1991game}
D.~Fudenberg and J.~Tirole, {\em Game theory}.
\newblock MIT press, 1991.

\bibitem{roth2000game}
A.~E. Roth, ``Game theory as a tool for market design,'' in {\em Game practice:
  Contributions from applied game theory}, pp.~7--18, Springer, 2000.

\bibitem{Aliasgari13}
M.~Aliasgari, M.~Blanton, Y.~Zhang, and A.~Steele, ``Secure computation on
  floating point numbers,'' in {\em Network \& Distributed System Security
  Symposium, {NDSS} 2013}, The Internet Society, 2013.

\bibitem{Kamm2015}
L.~Kamm and J.~Willemson, ``Secure floating point arithmetic and private
  satellite collision analysis,'' {\em International Journal of Information
  Security}, vol.~14, pp.~531--548, Nov 2015.

\bibitem{Catrina2010}
O.~Catrina and A.~Saxena, ``Secure computation with fixed-point numbers,'' in
  {\em Financial Cryptography and Data Security} (R.~Sion, ed.), (Berlin,
  Heidelberg), pp.~35--50, Springer Berlin Heidelberg, 2010.

\bibitem{Cock2015}
M.~d. Cock, R.~Dowsley, A.~C. Nascimento, and S.~C. Newman, ``Fast, privacy
  preserving linear regression over distributed datasets based on
  pre-distributed data,'' in {\em Proceedings of the 8th ACM Workshop on
  Artificial Intelligence and Security}, AISec '15, p.~3–14, ACM, 2015.

\bibitem{Gascon17}
A.~Gascón, P.~Schoppmann, B.~Balle, M.~Raykova, J.~Doerner, S.~Zahur, and
  D.~Evans, ``Privacy-preserving distributed linear regression on
  high-dimensional data,'' {\em Proceedings on Privacy Enhancing Technologies},
  vol.~2017, no.~4, pp.~345--364, 2017.

\bibitem{Feigenbaum2006}
J.~Feigenbaum, Y.~Ishai, T.~Malkin, K.~Nissim, M.~J. Strauss, and R.~N. Wright,
  ``Secure multiparty computation of approximations,'' {\em ACM Trans.
  Algorithms}, vol.~2, p.~435–472, jul 2006.

\bibitem{Ieee754}
{IEEE Comp. Soc.}, ``{IEEE} standard for floating-point arithmetic,'' {\em
  {IEEE} Std 754-2019 (Revision of {IEEE} 754-2008)}, pp.~1--84, 2019.

\bibitem{bfloat}
{Google Cloud}, ``The bfloat16 numerical format.''
\newblock Part of Cloud TPU Documentation.

\bibitem{Gustafson17}
J.~L. Gustafson and I.~T. Yonemoto, ``Beating floating point at its own game:
  Posit arithmetic,'' {\em Supercomputing frontiers and innovations}, vol.~4,
  no.~2, pp.~71--86, 2017.

\bibitem{Tjell21}
K.~{Tjell} and R.~{Wisniewski}, ``{Privacy in Distributed Computations based on
  Real Number Secret Sharing},'' {\em arXiv e-prints}, p.~arXiv:2107.00911,
  July 2021.

\bibitem{Pullonen15}
P.~Pullonen and S.~Siim, ``Combining secret sharing and garbled circuits for
  efficient private {IEEE} 754 floating-point computations,'' in {\em Financial
  Cryptography and Data Security} (M.~Brenner, N.~Christin, B.~Johnson, and
  K.~Rohloff, eds.), (Berlin, Heidelberg), pp.~172--183, Springer Berlin
  Heidelberg, 2015.

\bibitem{Dimitrov16}
V.~Dimitrov, L.~Kerik, T.~Krips, J.~Randmets, and J.~Willemson, ``Alternative
  implementations of secure real numbers,'' in {\em Proceedings of the 2016 ACM
  SIGSAC Conference on Computer and Communications Security}, CCS '16, (New
  York, NY, USA), p.~553–564, Association for Computing Machinery, 2016.

\bibitem{Liu16}
X.~Liu, R.~H. Deng, W.~Ding, R.~Lu, and B.~Qin, ``Privacy-preserving outsourced
  calculation on floating point numbers,'' {\em {IEEE} Transactions on
  Information Forensics and Security}, vol.~11, no.~11, pp.~2513--2527, 2016.

\bibitem{Kuskonmaz22}
B.~{Kuskonmaz}, J.~{Skovsted Gundersen}, and R.~{Wisniewski}, ``{Investigation
  of Alternative Measures for Mutual Information},'' {\em arXiv e-prints},
  p.~arXiv:2202.00956, Feb. 2022.

\bibitem{Dwork06}
C.~Dwork, F.~McSherry, K.~Nissim, and A.~Smith, ``Calibrating noise to
  sensitivity in private data analysis,'' in {\em Theory of Cryptography}
  (S.~Halevi and T.~Rabin, eds.), (Berlin, Heidelberg), pp.~265--284, Springer
  Berlin Heidelberg, 2006.

\bibitem{Dwork08}
C.~Dwork, ``Differential privacy: A survey of results,'' in {\em Theory and
  Applications of Models of Computation} (M.~Agrawal, D.~Du, Z.~Duan, and
  A.~Li, eds.), (Berlin, Heidelberg), pp.~1--19, Springer Berlin Heidelberg,
  2008.

\bibitem{Pentyala22}
S.~{Pentyala}, D.~{Railsback}, R.~{Maia}, R.~{Dowsley}, D.~{Melanson},
  A.~{Nascimento}, and M.~{De Cock}, ``{Training Differentially Private Models
  with Secure Multiparty Computation},'' {\em arXiv e-prints},
  p.~arXiv:2202.02625, Feb. 2022.

\bibitem{Gong20}
M.~Gong, Y.~Xie, K.~Pan, K.~Feng, and A.~Qin, ``A survey on differentially
  private machine learning [review article],'' {\em {IEEE} Computational
  Intelligence Magazine}, vol.~15, no.~2, pp.~49--64, 2020.

\bibitem{ali2020voluntary}
S.~N. Ali, G.~Lewis, and S.~Vasserman, ``Voluntary disclosure and personalized
  pricing,'' in {\em Proceedings of the 21st ACM Conference on Economics and
  Computation}, pp.~537--538, 2020.

\bibitem{jones2020nonrivalry}
C.~I. Jones and C.~Tonetti, ``Nonrivalry and the economics of data,'' {\em
  American Economic Review}, vol.~110, no.~9, pp.~2819--58, 2020.

\bibitem{cummings2015truthful}
R.~Cummings, S.~Ioannidis, and K.~Ligett, ``Truthful linear regression,'' in
  {\em Conference on Learning Theory}, pp.~448--483, PMLR, 2015.

\bibitem{feng2021uncovering}
Z.~Feng, J.~Chen, and Y.~Zhu, ``Uncovering value of correlated data: trading
  data based on iterative combinatorial auction,'' in {\em 2021 {IEEE} 18th
  International Conference on Mobile Ad Hoc and Smart Systems (MASS)},
  pp.~260--268, IEEE, 2021.

\bibitem{pandey2020crowdsourcing}
S.~R. Pandey, N.~H. Tran, M.~Bennis, Y.~K. Tun, A.~Manzoor, and C.~S. Hong, ``A
  crowdsourcing framework for on-device federated learning,'' {\em {IEEE}
  Transactions on Wireless Communications}, vol.~19, no.~5, pp.~3241--3256,
  2020.

\bibitem{Luo09}
W.~Luo and T.~Wang, ``Design and analysis of private-preserving dot product
  protocol,'' in {\em 2009 International Conference on Electronic Computer
  Technology. {ICECT} 2009}, (Los Alamitos, CA, USA), pp.~531--535, IEEE
  Computer Society, feb 2009.

\bibitem{Hu16}
C.~Hu, R.~Li, W.~Li, J.~Yu, Z.~Tian, and R.~Bie, ``Efficient privacy-preserving
  schemes for dot-product computation in mobile computing,'' in {\em
  Proceedings of the 1st ACM Workshop on Privacy-Aware Mobile Computing}, PAMCO
  '16, (New York, NY, USA), p.~51–59, Association for Computing Machinery,
  2016.

\bibitem{Canetti20}
R.~Canetti, ``Universally composable security,'' {\em J. ACM}, vol.~67, sep
  2020.

\bibitem{Cramer2015}
R.~Cramer, I.~B. Damgård, and J.~B. Nielsen, {\em Secure Multiparty
  Computation and Secret Sharing}.
\newblock Cambridge University Press, 2015.

\bibitem{Beaver91}
D.~Beaver, ``Efficient multiparty protocols using circuit randomization,'' in
  {\em Advances in Cryptology --- CRYPTO '91} (J.~Feigenbaum, ed.), (Berlin,
  Heidelberg), pp.~420--432, Springer Berlin Heidelberg, 1992.

\bibitem{Gilboa99}
N.~Gilboa, ``Two party {RSA} key generation,'' in {\em Advances in Cryptology
  --- CRYPTO' 99} (M.~Wiener, ed.), (Berlin, Heidelberg), pp.~116--129,
  Springer Berlin Heidelberg, 1999.

\bibitem{MASCOT}
M.~Keller, E.~Orsini, and P.~Scholl, ``{MASCOT}: Faster malicious arithmetic
  secure computation with oblivious transfer,'' in {\em Proceedings of the 2016
  ACM SIGSAC Conference on Computer and Communications Security}, CCS '16, (New
  York, NY, USA), p.~830–842, Association for Computing Machinery, 2016.

\bibitem{Keller18}
M.~Keller, V.~Pastro, and D.~Rotaru, ``Overdrive: Making {SPDZ} great again,''
  in {\em Advances in Cryptology -- EUROCRYPT 2018} (J.~B. Nielsen and
  V.~Rijmen, eds.), (Cham), pp.~158--189, Springer International Publishing,
  2018.

\bibitem{Boyle20}
E.~Boyle, G.~Couteau, N.~Gilboa, Y.~Ishai, L.~Kohl, and P.~Scholl, ``Efficient
  pseudorandom correlation generators from ring-{LPN},'' in {\em Advances in
  Cryptology -- CRYPTO 2020} (D.~Micciancio and T.~Ristenpart, eds.), (Cham),
  pp.~387--416, Springer International Publishing, 2020.

\bibitem{Hotelling53}
H.~Hotelling, ``New light on the correlation coefficient and its transforms,''
  {\em J. R. Stat. Soc. B}, vol.~15, no.~2, pp.~193--232, 1953.

\bibitem{CoverThomas}
T.~M. Cover and J.~A. Thomas, {\em Elements of Information Theory}.
\newblock John Wiley \& Sons, 2~ed., 2005.

\end{thebibliography}

\end{document}